\documentclass[12pt,reqno]{amsart}
\usepackage{graphicx}
\usepackage{amscd}
\usepackage{amsmath}
\usepackage{epsfig}
\usepackage{amsfonts}
\usepackage{amssymb}

\providecommand{\U}[1]{\protect\rule{.1in}{.1in}}
\providecommand{\U}[1]{\protect\rule{.1in}{.1in}}
\allowdisplaybreaks

\theoremstyle{plain}

\numberwithin{equation}{section}

\input{tcilatex}

\begin{document}
%
%
%
%
%
%
%
%
\title[ Berry Phase for Simple Harmonic Oscillators]{The Berry Phase for
Simple Harmonic Oscillators}
\author{Sergei K. Suslov}
\address{School of Mathematical and Statistical Sciences \& Mathematical,
Computational and Modeling Sciences Center, Arizona State University, Tempe,
AZ 85287--1804, U.S.A.}
\email{sks@asu.edu}
\urladdr{http://hahn.la.asu.edu/\symbol{126}suslov/index.html}

\begin{abstract}
We evaluate the Berry phase for a \textquotedblleft
missing\textquotedblright\ family of the square integrable wavefunctions for
the linear harmonic oscillator, which cannot be derived by the separation of
variables (in a natural way). Instead, it is obtained by the action of the
maximal kinematical\ invariance group on the standard solutions. A simple
closed formula for the phase (in terms of elementary functions) is found by
integration with the help of a computer algebra system.
\end{abstract}

\maketitle

Recent reports on observations of the dynamical Casimir effect \cite%
{Lahetal11}, \cite{Wilsonetal11} strengthens the interest to `nonclassical'
states in quantum optics and generalized harmonic oscillators \cite%
{Dodonov02}, \cite{Dodonov10}, \cite{Dod:Mal:Man75}, \cite{Dod:Man79}, \cite%
{Dodonov:Man'koFIAN87}, \cite{Dodonov:Man'ko03}, \cite{Malkin:Man'ko79}, \cite{Man'koCasimir}
and \cite{Nationetal12}.
The amplification of quantum fluctuations by modulating parameters of an oscillator is
closely related to the process of particle production in quantum fields \cite{Dodonov10},
\cite{Jacobson04}, \cite{Man'koCasimir} and \cite{Nationetal12}.
Other dynamical amplification mechanisms include the Unruh effect \cite{Unruh76} and Hawking
radiation \cite{Hawking74}, \cite{Hawking75}.

The purpose of this paper is to evaluate the Berry phase for certain
\textquotedblleft missing\textquotedblright\ solutions of the time-dependent
Schr\"{o}dinger equation for the linear harmonic oscillator as an instructive example.
Applications will be discussed elsewhere.

\section{Hidden Solutions}

The time-dependent Schr\"{o}dinger equation for the simple harmonic
oscillator,%
\begin{equation}
2i\psi _{t}+\psi _{xx}-x^{2}\psi =0,  \label{Schroudinger}
\end{equation}%
has the following six-parameter family of (square integrable) solutions \cite%
{LopSusVegaHarm}:
\begin{equation}
\psi _{n}\left( x,t\right) =\frac{e^{i\left( \alpha \left( t\right)
x^{2}+\delta \left( t\right) x+\kappa \left( t\right) \right) +i\left(
2n+1\right) \gamma \left( t\right) }}{\sqrt{2^{n}n!\mu \left( t\right) \sqrt{%
\pi }}}\ e^{-\left( \beta \left( t\right) x+\varepsilon \left( t\right)
\right) ^{2}/2}\ H_{n}\left( \beta \left( t\right) x+\varepsilon \left(
t\right) \right) ,  \label{WaveFunction}
\end{equation}%
where $H_{n}\left( x\right) $ are the Hermite polynomials \cite{Ni:Su:Uv} and%
\begin{eqnarray}
\mu \left( t\right) &=&\mu _{0}\sqrt{\beta _{0}^{4}\sin ^{2}t+\left( 2\alpha
_{0}\sin t+\cos t\right) ^{2}},  \label{hhM} \\
\alpha \left( t\right) &=&\frac{\alpha _{0}\cos 2t+\sin 2t\ \left( \beta
_{0}^{4}+4\alpha _{0}^{2}-1\right) /4}{\beta _{0}^{4}\sin ^{2}t+\left(
2\alpha _{0}\sin t+\cos t\right) ^{2}},  \label{hhA} \\
\beta \left( t\right) &=&\frac{\beta _{0}}{\sqrt{\beta _{0}^{4}\sin
^{2}t+\left( 2\alpha _{0}\sin t+\cos t\right) ^{2}}},  \label{hhB} \\
\gamma \left( t\right) &=&\gamma _{0}-\frac{1}{2}\arctan \frac{\beta
_{0}^{2}\sin t}{2\alpha _{0}\sin t+\cos t},  \label{hhG} \\
\delta \left( t\right) &=&\frac{\delta _{0}\left( 2\alpha _{0}\sin t+\cos
t\right) +\varepsilon _{0}\beta _{0}^{3}\sin t}{\beta _{0}^{4}\sin
^{2}t+\left( 2\alpha _{0}\sin t+\cos t\right) ^{2}},  \label{hhD} \\
\varepsilon \left( t\right) &=&\frac{\varepsilon _{0}\left( 2\alpha _{0}\sin
t+\cos t\right) -\beta _{0}\delta _{0}\sin t}{\sqrt{\beta _{0}^{4}\sin
^{2}t+\left( 2\alpha _{0}\sin t+\cos t\right) ^{2}}},  \label{hhE} \\
\kappa \left( t\right) &=&\kappa _{0}+\sin ^{2}t\ \frac{\varepsilon
_{0}\beta _{0}^{2}\left( \alpha _{0}\varepsilon _{0}-\beta _{0}\delta
_{0}\right) -\alpha _{0}\delta _{0}^{2}}{\beta _{0}^{4}\sin ^{2}t+\left(
2\alpha _{0}\sin t+\cos t\right) ^{2}}  \label{hhK} \\
&&+\frac{1}{4}\sin 2t\ \frac{\varepsilon _{0}^{2}\beta _{0}^{2}-\delta
_{0}^{2}}{\beta _{0}^{4}\sin ^{2}t+\left( 2\alpha _{0}\sin t+\cos t\right)
^{2}}  \notag
\end{eqnarray}%
($\mu _{0}\neq 0,$ $\alpha _{0},$ $\beta _{0}\neq 0,$ $\gamma _{0},$ $\delta
_{0},$ $\varepsilon _{0},$ $\kappa _{0}$ are real initial data). These
solutions\ have been derived analytically in the framework of a unified
approach to generalized harmonic oscillators (see, for example, \cite%
{Cor-Sot:Lop:Sua:Sus}, \cite{Cor-Sot:Sua:SusInv}, \cite{Lan:Lop:Sus},
\cite{Wolf81}, \cite{Yeon:Lee:Um:George:Pandey93}
and the
references therein). They are also verified by a direct substitution with
the aid of {\sl{Mathematica}} computer algebra system \cite{Kouchan11},
\cite{Lop:Sus:VegaMath}. (The\ simplest special case $\mu _{0}=\beta _{0}=1$
and $\alpha _{0}=\gamma _{0}=\delta _{0}=\varepsilon _{0}=\kappa _{0}=0$
reproduces the textbook solution obtained by the separation of variables
\cite{Schroedinger}, \cite{Flu}, \cite{La:Lif}, \cite{Merz}.
The shape-preserving oscillator evolutions occur when $\alpha_0=0$ and $%
\beta_0=1$ and a special case when $\alpha _{0}=0$ is
discussed in \cite{Husimi53}.
More details on the derivation of these formulas and some
{\sl{Mathematica}} animations, revealing a new feature -- an oscillation
in space of the probability density $\left\vert \psi \left( x,t\right)
\right\vert ^{2}$ -- of these solutions, can be found in Refs.~\cite%
{Kouchan11}, \cite{Lop:Sus:VegaGroup} and \cite{Lop:Sus:VegaMath}.)

The \textquotedblleft dynamic harmonic oscillator states\textquotedblright\ (%
\ref{WaveFunction})--(\ref{hhK}) are eigenfunctions,%
\begin{equation}
E\left( t\right) \psi _{n}\left( x,t\right) =\left( n+\frac{1}{2}\right)
\psi _{n}\left( x,t\right) ,  \label{EigenValueProblem}
\end{equation}%
of the time-dependent quadratic invariant,%
\begin{equation}
E\left( t\right) =\frac{1}{2}\left[ \frac{\left( p-2\alpha x-\delta \right)
^{2}}{\beta ^{2}}+\left( \beta x+\varepsilon \right) ^{2}\right] ,\qquad
\frac{d}{dt}\langle E\rangle =0,  \label{QuadraticInvariant}
\end{equation}%
where $p=i^{-1}\partial /\partial x$ and the required operator identity,%
\begin{equation}
\frac{\partial E}{\partial t}+i^{-1}\left[ E,H\right] =0,\qquad H=\frac{1}{2}%
\left( p^{2}+x^{2}\right) ,  \label{InvariantDer}
\end{equation}%
holds \cite{SanSusVin}.

The (isomorphic) maximum kinematical invariance groups of the free particle
and harmonic oscillator were introduced in \cite{AndersonPlus72}, \cite%
{AndersonII72}, \cite{Hagen72}, \cite{JACKIW80}, \cite{Niederer72} and \cite%
{Niederer73} (see also \cite{BoySharpWint}, \cite{KalninsMiller74}, \cite%
{Miller77}, \cite{VinetZhedanov2011} and the references therein). We have
established a connection with certain Ermakov-type system which allows us to
bypass a complexity of the traditional Lie algebra approach \cite%
{Lop:Sus:VegaGroup}, \cite{LopSusVegaHarm}. (A general procedure of
obtaining new solutions by acting on any set of given ones by enveloping
algebra of generators of the Heisenberg--Weyl group is described in \cite%
{Dodonov:Man'koFIAN87}; see also \cite{Bag:Bel:Ter83}, \cite%
{Belov:Karavaev1987} and \cite{Dod:Man79}.)

\section{Evaluation of the Phase}

The holonomic effect in quantum mechanics known as Berry's phase \cite%
{Berry84}, \cite{Berry85}, \cite{Simon83}, \cite%
{WilczekZee84} has received considerable attention over the years
(see, for example, \cite{Dodonov:Man'koBerry88}, \cite{Vinitskietal90} and
the other references in \cite{SanSusVin}). The derivative of Berry's phase has
been recently calculated for the generalized harmonic oscillators as follows
\cite{SanSusVin}:%
\begin{equation}
\frac{d\theta _{n}}{dt}=-\beta ^{-2}\left( \varepsilon ^{2}+n+\frac{1}{2}%
\right) \frac{d\alpha }{dt}+\varepsilon \beta ^{-1}\frac{d\delta }{dt}-\frac{%
d\kappa }{dt},  \label{BerryPhase}
\end{equation}%
where we are going to use (\ref{hhA})--(\ref{hhK}) and simplify. Integrating
by parts, one gets
\begin{eqnarray}
\theta _{n} &=&-\left( n+\frac{1}{2}\right) \int \beta ^{-2}\frac{d\alpha }{%
dt}\ dt-\left( \frac{\varepsilon }{\beta }\right) ^{2}\alpha +\frac{%
\varepsilon \delta }{\beta }-\kappa   \label{BerryIntegral} \\
&&+\int \left[ \alpha \frac{d}{dt}\left( \frac{\varepsilon }{\beta }\right)
^{2}-\delta \frac{d}{dt}\left( \frac{\varepsilon }{\beta }\right) \right] \
dt.  \notag
\end{eqnarray}%
Here,%
\begin{eqnarray}
&&4\beta _{0}^{2}\left[ \left( \frac{\varepsilon }{\beta }\right) ^{2}\alpha
-\frac{\varepsilon \delta }{\beta }+\kappa \right] =2\beta _{0}\left( 2\beta
_{0}\kappa _{0}-\delta _{0}\varepsilon _{0}\right)   \label{ConstantTerm} \\
&&\qquad +2\varepsilon _{0}\left( 2\alpha _{0}\varepsilon _{0}-\beta
_{0}\delta _{0}\right) \cos 2t+\left[ \left( 2\alpha _{0}\varepsilon
_{0}-\beta _{0}\delta _{0}\right) ^{2}-\varepsilon _{0}^{2}\right] \sin 2t,
\notag
\end{eqnarray}%
\begin{eqnarray}
&&4\beta _{0}^{2}\int \left[ \alpha \frac{d}{dt}\left( \frac{\varepsilon }{%
\beta }\right) ^{2}-\delta \frac{d}{dt}\left( \frac{\varepsilon }{\beta }%
\right) \right] \ dt=2t\left[ \left( 2\alpha _{0}\varepsilon _{0}-\beta
_{0}\delta _{0}\right) ^{2}+\varepsilon _{0}^{2}\right]
\label{FreeIntegral} \\
&&\qquad +2\varepsilon _{0}\left( 2\alpha _{0}\varepsilon _{0}-\beta
_{0}\delta _{0}\right) \cos 2t+\left[ \left( 2\alpha _{0}\varepsilon
_{0}-\beta _{0}\delta _{0}\right) ^{2}-\varepsilon _{0}^{2}\right] \sin 2t,
\notag
\end{eqnarray}%
\begin{equation}
\int \beta ^{-2}\frac{d\alpha }{dt}\ dt=-t\frac{4\alpha _{0}^{2}+\beta
_{0}^{4}+1}{2\beta _{0}^{2}}+\arctan \left[ \frac{2\alpha _{0}+\left(
4\alpha _{0}^{2}+\beta _{0}^{4}\right) \tan t}{\beta _{0}^{2}}\right]
\label{nIntegral}
\end{equation}%
with the aid of {\sl{Mathematica}} (the notebook is available from the
author's website \cite{SuslovMath}).

Finally, we evaluate Berry's phase in a closed form:%
\begin{eqnarray}
\theta _{n}\left( t\right)  &=&-\left( n+\frac{1}{2}\right) \left[ \arctan
\left( \frac{2\alpha _{0}+\left( 4\alpha _{0}^{2}+\beta _{0}^{4}\right) \tan
t}{\beta _{0}^{2}}\right) -\arctan \left( \frac{2\alpha _{0}}{\beta _{0}^{2}}%
\right) -t\frac{4\alpha _{0}^{2}+\beta _{0}^{4}+1}{2\beta _{0}^{2}}\right]
\notag \\
&&+t\frac{\left( 2\alpha _{0}\varepsilon _{0}-\beta _{0}\delta _{0}\right)
^{2}+\varepsilon _{0}^{2}}{2\beta _{0}^{2}},\qquad \theta _{n}\left(
0\right) =0.  \label{PhaseFinal}
\end{eqnarray}%
(This expression has been verified by differentiation with the help of
{\sl{Mathematica}} once again \cite{SuslovMath}. Examples are presented in
Figure~1.) To the best of our knowledge, this formula\ is also missing in
the available literature --- in the\ simplest case $\beta _{0}=1$ and $%
\alpha _{0}=\gamma _{0}=\delta _{0}=\varepsilon _{0}=\kappa _{0}=0,$ one
obtains $\theta _{n}=0,$ which is a well-known result for the textbook
solutions. Our formula implies that for the shape-preserving oscillator
evolutions, when $\alpha _{0}=0$ and $\beta _{0}=1$, the phase does not
depend on $n.$

On the second hand, Eq.~(42) of Ref.~\cite{SanSusVin} gives an alternative
formula for evaluation of the phase,%
\begin{equation}
\theta _{n}=\left( 2n+1\right) \gamma +\int \langle H\rangle \ dt,\label%
{ThetaAnother}
\end{equation}%
where%
\begin{equation}
\langle H\rangle =\frac{1}{2}\left[ \langle p^{2}\rangle +\langle
x^{2}\rangle \right] =\left( n+\frac{1}{2}\right) \dfrac{1+4\alpha
_{0}^{2}+\beta _{0}^{4}}{2\beta _{0}^{2}}+\frac{\left( 2\alpha
_{0}\varepsilon _{0}-\beta _{0}\delta _{0}\right) ^{2}+\varepsilon _{0}^{2}}{%
2\beta _{0}^{2}}\label{Energy}
\end{equation}%
by (A.3)--(A.5) of Ref.\ \cite{LopSusVegaHarm}. As a result one gets%
\begin{equation}
\theta _{n}=-\left( n+\frac{1}{2}\right) \arctan \frac{\beta _{0}\tan t}{%
1+2\alpha _{0}\tan t}+\left( n+\frac{1}{2}\right) \dfrac{1+4\alpha
_{0}^{2}+\beta _{0}^{4}}{2\beta _{0}^{2}}t+\frac{\left( 2\alpha
_{0}\varepsilon _{0}-\beta _{0}\delta _{0}\right) ^{2}+\varepsilon _{0}^{2}}{%
2\beta _{0}^{2}}t,\label{THETA}
\end{equation}%
which is equivalent to our previuos expression (\ref{PhaseFinal}) up to
an elementary transformation.

\section{A Conclusion}

In addition to the oscillation in space of the probability density $%
\left\vert \psi \left( x,t\right) \right\vert ^{2},$ which has already been
computer animated in \cite{Lop:Sus:VegaMath} and \cite{LopSusVegaHarm}, the
\textquotedblleft dynamic harmonic states\textquotedblright\ (\ref%
{WaveFunction})--(\ref{hhK}) possess the nontrivial Berry phase. These two
distinguished features of the quantum motion under consideration might be
observed in a clever experiment.

Moreover, the electromagnetic field quantization presents the EM field
in nonstationary media as a set of harmonic oscillators \cite{Dodonov10}
and \cite{Dod:Klim:Nik93}.
Thus the Berry phase evaluated in this paper is somehow related to
the squeezed states of light
which are produced in the process of parametric amplification.
(See also Ref.~\cite{Xiaoetal10} for other possible applications.)


%
\begin{figure}[htbp]
\centering \scalebox{1.25}{\includegraphics{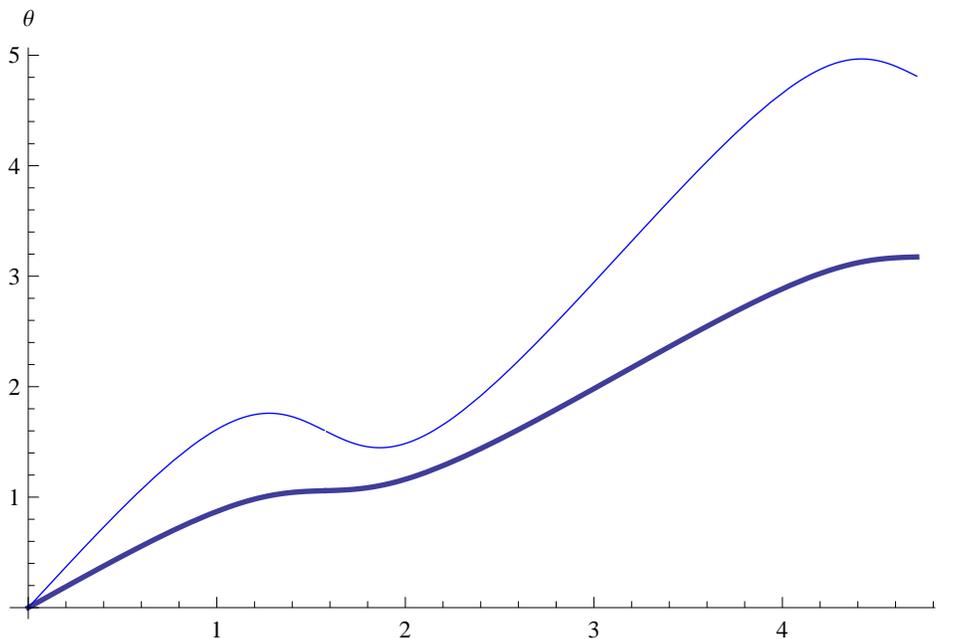}}
\caption{The phases $\theta_0\/\left( t\right) $ and
$\theta_1\/\left( t\right) $
with $\alpha_0=\gamma_0=\varepsilon_0=0$, $\beta_0=2/3$ and $\delta_0=1$
(thick and thin lines, respectively).}
\end{figure}
%

%
%

\noindent \textbf{Acknowledgments.\/} We thank Michael Berry, Andrew
Bremner, Carlos Castillo-Ch\'{a}vez, Victor V. Dodonov, Chris\-toph
Koutschan, Elliott Lieb, Francisco F.~L\'{o}pez-Ruiz, Vladimir I. Man'ko,
Benjamin R. Morin, Sergey I. Kryuchkov, Andreas Ruffing, Vladimir M.
Shabaev, Barbara Sanborn, Luc Vinet and Doron Zeilberger for support,
valuable discussions and encouragement. This paper is written as a result of
author's visit to RISC, Research Institute for Symbolic Computation, and The
Erwin Schr\"{o}dinger International Institute for Mathematical Physics ---
we thank Peter Paule, RISC, Johannes Kepler Universit\"{a}t Linz, and
Christian Krattenthaler, Fakult\"{a}t f\"{u}r Mathematik, Universit\"{a}t
Wien, for their hospitality. We are grateful to Catherine Boucher of
Wolfram Science Group for an independent {\sl{Mathematica}} verification of the
\textquotedblleft missing\textquotedblright\ solutions.

\end{document}